\documentclass[12pt]{article}
\usepackage{epsfig}

\begin{document}

\begin{center}
\bf{\large LABORATORY SIMULATION OF ASTROPHYSICAL JETS WITHIN FACILITIES  
OF PLASMA FOCUS TYPE}
\end{center}

\begin{center}
{VYACHESLAV I. KRAUZ}
\end{center}

\begin{center}
{\it NRC Kurchatov Institute, Kurchatov Sq. 1,  Moscow, 123182, Russia \\
Moscow Institute of Physics and Technology, Institutsky per., 9,  \\
Dolgoprudny, 141700, Russia  \\                                  
Krauz\_VI@nrcki.ru}
\end{center}

\begin{center}
{VASILY S. BESKIN}
\end{center}

\begin{center}
{\it Lebedev Physical Institute, Leninsky prosp., 53,
Moscow, 119991, Russia\\
Moscow Institute of Physics and Technology, Institutsky per., 9, \\ 
Dolgoprudny, 141700, Russia  \\
beskin@lpi.ru}
\end{center}

\begin{center}
{EVGENY P. VELIKHOV}
\end{center}

\begin{center}
{\it NRC Kurchatov Institute, Kurchatov Sq. 1,  Moscow, 123182, Russia \\
Moscow Institute of Physics and Technology, Institutsky per., 9, \\
Dolgoprudny, 141700, Russia \\
velikhov@mac.com} 
\end{center}


\begin{abstract}
A laboratory simulation of astrophysical processes is one of the intensively developed areas 
of plasma physics. A new series of experiments has been launched recently on the Plasma 
Focus type facility in NRC Kurchatov Institute. The main goal is to study the mechanisms 
of the jet stabilization, due to which it can propagate at distances much greater than 
their transverse dimensions. The experiments with stationary gas filling revealed regimes 
in which a narrowly collimated plasma jet was formed, the head of which was no wider than 
several centimeters at jet propagation distances of up to 100 cm. The PF-1000 (IFPiLM, 
Warsaw, Poland) and KPF-4 (SFTI, Sukhum, Abkhazia) experiments are aimed at creating profiled
initial gas distributions to control the conditions of plasma jet propagation in the ambient
plasma. Estimations of the dimensionless parameters, i.e. the Mach, Reynolds, and Peclet numbers
which were achieved during the experiments, showed that the PF-facilities can be used for the YSO
jets modelling. The future ex\-pe\-ri\-ments, which can allow one to understand the nature of the stable
plasma ejections observed in many astrophysical sources, are discussed. 
\end{abstract}



\section{Introduction }    

Laboratory simulation of astrophysical processes is one of the 
intensively developed areas of plasma physics. The obvious complexity 
of the theoretical simulation of the astrophysical objects is related 
to the absence of the targeted experiments, which allow one to observe 
the system response by varying the physical parameters. In recent decades, 
such experiments have been replaced by numerical simulation which yielded 
important results. At the same time, a lot of interesting astrophysical 
phenomena, including nonrelativistic ejections from the young stars, 
can be simulated in a laboratory experiment with observation of certain 
similarity laws~\cite{1, 2}. Such an approach is reasonable since the MHD 
equations, which govern both the astrophysical plasma jets and the 
laboratory plasma flows, permit the spatial and temporal scaling.

Considerable progress in simulating the astrophysical processes has been 
achieved in recent decades due to the appearance of a whole group of new 
facilities with high energy density, which were developed within the 
framework of the program of inertial controlled fusion, in particular, 
the modern Z-pinch systems~\cite{3} and high-power lasers~\cite{4}. The obvious 
advantages of laboratory modeling are the ability to actively control the 
parameters of the experiment, high repeatability, and well-developed diagnostic 
base. In addition, it is possible to model the dynamics of the process and 
phenomena that are not available for direct observation. In particular, 
interesting results were obtained on a high-power laser in the LULI laboratory 
 (Ecole Polytechnic), where it was shown that the superimposition of an 
external poloidal magnetic field can provide effective collimation of the 
plasma flow~\cite{5}. There are well-known studies of recent years on the Z-pinch 
installation MAGPIE (Imperial College London, Great Britain), which simulated 
some possible mechanisms of jet formation, the interaction of a supersonic, 
radiatively cooled plasma jet with an ambient medium and a number of other 
effects~\cite{6, 7}.

Plasma Focus (PF) is a device, which operating principle is also based 
on the Z-pinch effect. At the same time, installations of this type have a number 
of preferences, which allow us to arrange the original experiments aimed at 
modeling the plasma outflows from young stellar objects (YSO). In this paper, we 
substantiate the applicability of these facilities for such modeling, list the 
main results obtained and formulate a program for further research.

\section{Plasma Focus}

\begin{figure}[pb]
\centerline{\epsfig{file=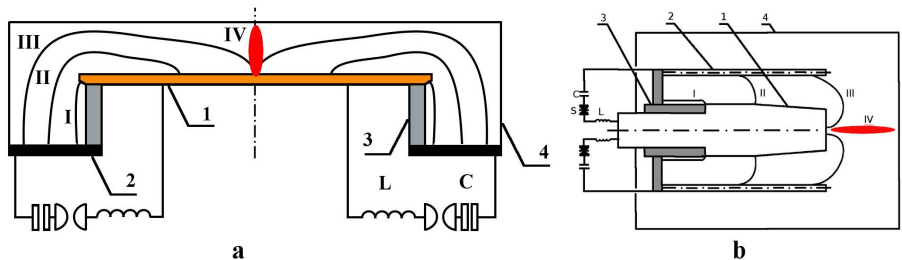,width=12cm}}
\vspace*{8pt}
\caption{
Principal scheme of plasma focus operation: a)  Filippov-type ($ D/L \geq 1$), 
b) Mather-type ($ D/L \ll 1$):
1 --- anode, 2 --- cathode, 3 --- insulator, 4 --- vacuum chamber, ' --- power supply, 
L --- external inductance,  S --- spark gap, I --- break-down phase; II --- run-down phase; 
III --- dense plasma focus phase, IV --- supersonic plasma outflow
}
\label{fig1}
\end{figure}

PF is well known as a source of intense plasma flows, which are widely used 
in various fields of science and technology, such as radiation material science, 
modification of materials, etc. Plasma flow is formed in the stage of pinching, 
which lasts several hundred nanoseconds, in the pinch area with a diameter of approx. 1 cm 
and a length of (3-5) cm. Two most popular PF systems are known, which differ in the 
geometry of the discharge system: Filippov-type~\cite{8} and Mather-type~\cite{9} 
systems. The principal scheme of their operation is shown on Fig.~\ref{fig1}.  

After the preliminary pump-out, the chamber is filled with working gas (hydrogen, 
deuterium, helium, neon, argon, and their mixtures depending on the formulated tasks) 
under a pressure of a few Torr. When the spark gap switches on, a high voltage of power 
supply (condenser bank) is applied between the anode and the cathode, which leads to a 
breakdown of the working gas. The resulting plasma current-carring sheath (PCS) moves 
under the action of the Amp\`ere force toward the discharge system axis, where the plasma 
pinch occurs. The pinching is accompanied by a drop in the discharge current, and 
accordingly the appearance of a sharp dip in its derivative.  

In our opinion, the scheme of the experiment with PF has a number of advantages. 
First of all, there is the possibility to investigate the dynamics of the flow 
propagation in the ambient plasma at sufficiently large distances, more than two 
orders of magnitude greater than its initial transverse size. The presence of the 
ambient plasma in itself is an important factor that allows us simulating the interaction 
of the flow with the external environment. Moreover, the use of both stationary filling 
the working gas into the vacuum chamber and mode with gas-puffing and their combinations 
provides ample opportunities for the creation of various profiled gas distributions. 
This allows us modeling the density contrast effect (the ratio of the flow density to the density 
of the ambient plasma) on the jet dynamics. The use of various working gases, including 
strongly radiating, allows us investigating the role of radiation cooling in collimation 
and confinement of the jet. Finally, sufficiently large spatial dimensions of the flow 
(in comparison with the laser and Z-pinches experiment) made it possible to use magnetic 
probe techniques and allowed to investigate the distribution of magnetic fields.  

We deliberately did not concentrate on the mechanisms of flow generation, because we 
understand that they can differ significantly from those in astrophysics. However, the 
formed jet propagates then in the interstellar gas, and it is quite accessible for the 
simulation. Therefore the main our goal is to study the mechanisms of the jet stabilization, 
due to which it can propagate at distances much greater than its transverse dimension.  

\begin{figure}[pb]
\centerline{\epsfig{file=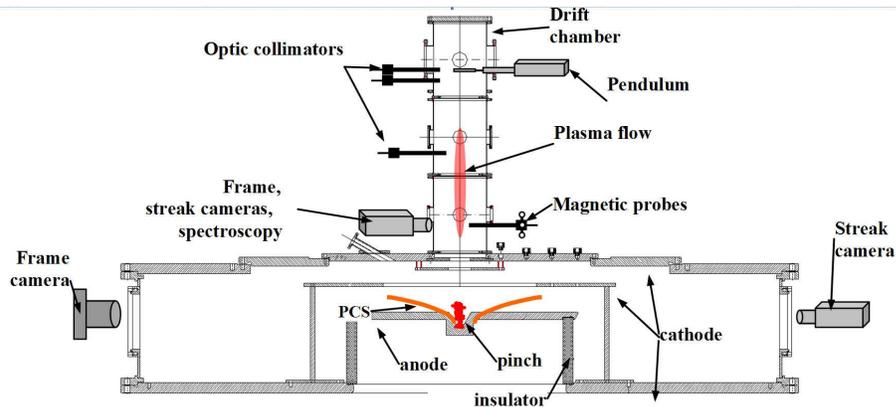,width=12cm}}
\vspace*{8pt}
\caption{Scheme of the PF-3 experiment}
\label{fig2}
\end{figure}

A new series of experiments devoted to the astrophysical jets simulation has been launched
recently on the PF-3 facility in NRC Kurchatov Institute~\cite{10}. The key difference of 
our experiments was the possibility of studying the process of propagation of jet plasma flows. 
For this PF-3 facility has been upgraded. A new three-section diagnostic drift chamber was
designed, which allowed one to measure the jet and the ambient plasma parameters at distances 
of up to 100 cm from the plane of the anode, which was conventionally assumed to be the 
flow generation region (see Fig.~\ref{fig2})~\cite{11}. 

\begin{figure}[pb]
\centerline{\epsfig{file=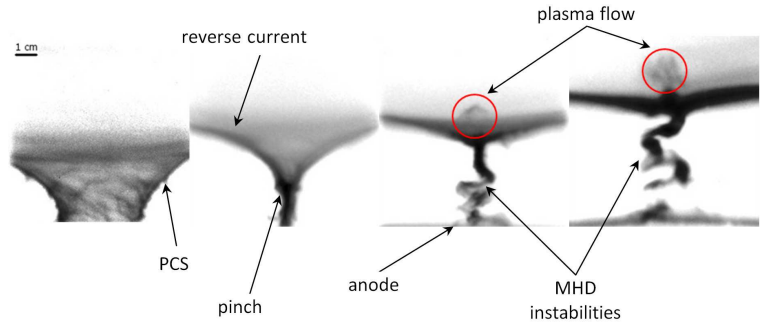,width=10cm}}
\vspace*{8pt}
\caption{Plasma-flow formation (from left to right): the stage of convergence of the
plasma-current sheath to the axis, the pinch stage and the stages of the plasma flow 
formation. The pictures were obtained with optical frame cameras, exposure time is 12 ns}
\label{fig3}
\end{figure}

It was found that the compact plasma jets moving along the axis occur at the stage of 
the pinch decay and developing the MHD instabilities (Fig.\ref{fig3})\cite{11, 12, 13}. 
The initial jet velocity, determined from a series of successive snapshots in a frame mode, 
$V_{0} \geq 10^7$ cm/s, regardless of the operating gas used.  This velocity is close 
to the velocity of plasma outflows from young stars and exceeds the velocity of the 
PCS lifting in the axial direction under the action of Amp\`ere forces. After a certain 
moment of time, the flow breaks away from the main current sheath and from the pinch 
and begins to live their lives, regardless of them. 

\begin{figure}[pb]
\centerline{\epsfig{file=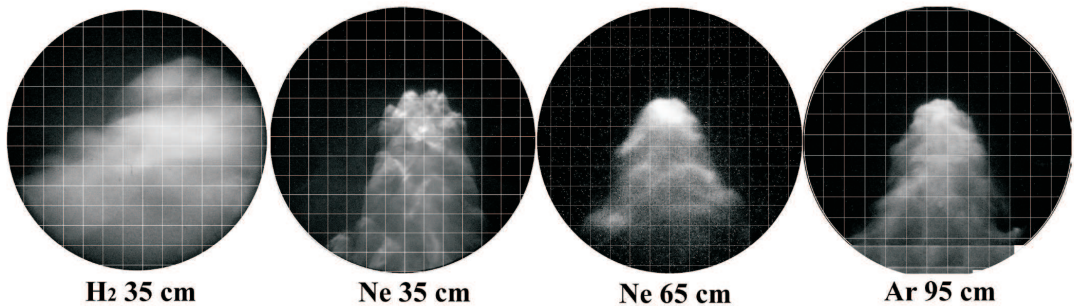,width=15cm}}
\vspace*{8pt}
\caption{Frame camera pictures of plasma flow front at different distances from the anode 
plane and at operation with different gases. Scale is 1 cm}
\label{fig4}
\end{figure}

One of the main obtained results is the finding the regimes with the formation of plasma 
object, preserving its compact size when spread over large distances\cite{11, 14}. The 
transverse dimension of the flow head does not exceed a few cm at propagation distances 
more than 100 cm (Fig.~\ref{fig4})~\cite{13, 15}.  This implies that there are mechanisms 
for stabilizing/confining the plasma flow, which was the subject of our further research. 

First of all, it was necessary to make sure that the streams generated in the PF discharge 
could be used for modeling the jets from YSO. As is known, the key properties of astrophysical 
outflows are the presence of a regular longitudinal magnetic field, longitudinal electric 
current, as well as axial rotation. Finally, it was necessary to show the correspondence 
of the basic dimensionless parameters to the known scaling relations.  To this end, a 
series of studies was carried out, including measurements of plasma parameters, flow 
velocity, magnetic fields, etc. 

Measurements of density and temperature were performed using the spectral technique~\cite{16, 17, 18}.
In the experiments with helium and neon, the flow density determined by the Stark broadening 
at the distance of 35 cm from the anode plane (the first section of the drift chamber) was 
$(2-4) \times 10^{17}$ cm$^{-3}$, the electron plasma temperature of the jet was $(2-8)$ eV. 
The  concentration of the ambient plasma was $\sim (2-4) \times 10^{16}$ cm$^{-3}$. At a 
distance of 65 cm from the pinch, the concentration of the ambient neon plasma was outside 
the limits of spectral equipment registration: $N_{\rm i} < 10^{16}$ cm$^{-3}$. The maximum
electron concentration in a jet is \mbox{$(0.5-2) \times 10^{17}$ cm$^{-3}$,} which is lower than
at a distance of 35 cm. However, the density reduction can be associated not only with the
divergence of the flow, but also with a decrease in the degree of ionization due to plasma cooling
(the electron temperature of the plasma jet at a distance of 65 cm is $T_{\rm e} \sim$ 1 eV). The
latter conclusion is confirmed  by data from frame cameras, showing the weak divergence of the flow
(Fig.~\ref{fig4}). 

As was already noted, the initial flow velocity in the laboratory experiment was similar to 
the velocity observed in the plasma outflows from YSO. The flow dynamics in different 
gases were studied using optical collimators, which record the radiation from a small 
solid angle along the diameter of the chamber~\cite{11, 19}. The presence of the ambient plasma 
largely determines the evolution of the plasma jet when it moving along the system axis. 
The change in the flow velocity can be approximated by an expression $V = V_{0}\exp(-l/l_{0})$, 
where $V_{0}$ is the initial velocity, $l_{0}$ is the braking length of the flow.  It was shown 
that $l_{0}$ strongly depends on the working gas. At the same time, the initial velocity 
weakly depends on the type of the gas and corresponds well to the velocity determined 
by the frame cameras at the stage of the jet formation. 

Finally, with the help of magnetic probes, the distribution of magnetic fields has been 
studied. Several modifications of magnetic probes are made for measurements: N-channel
magnetic probe for measurements the radial distribution of the azimuthal component of the 
magnetic field and 4-channel ($B_{z} , B_{r} , B_{\varphi}$, optic) probe for
measurements of three components of the magnetic fields and  optical radiation of plasma 
(with photomultiplier). It was shown that the plasma flow moves with the frozen magnetic 
field of the order $(1-10)$ kG. Obtained radial distribution can be explained by the axial 
current in $(1-10)$ kA flowing in the zone near the axis with the radius of $(1-1.5)$ cm. 

The estimations of key dimensionless parameters such as Mach number, Reynolds 
(both hydrodynamic and magnetic) and Peclet number  made on the basis of the obtained 
data (Table 1) showed that the PF- type facilities really can be used to simulate the 
jets from YSO. This allowed us to proceed to the study of other properties of the outflows, 
which can be modelled with the help of PF installations. 

\begin{table}[ph]
\caption{Key dimensionless parameters}
\vspace{0.5cm}
\centering
\begin{tabular}{|c|c|c|c|}
\hline
& YSO &  & PF-3 \\
& &  & (35 cm above \\
&  &  & the anode) \\
\hline
Peclet & $10^{11} $ & $> 1$, convective & $> 10^{7}$  \\
 &  & heat transfer &  \\
\hline
Reynolds & $10^{13}$ & $\gg 1$, the & $10^{4}-10^{5}$ \\
&  & viscosity &  \\
&  & is important  & \\
\hline
Magnetic  & $10^{15}$ & $> 1$, magnetic & $\sim 100$ \\
Reynolds &  & field is frozen &  \\
\hline
Mach   & $10-50$ & $> 1$, the jet is & $> 10$ (for Ne \\
$(V_{\rm jet}/V_{\rm cs})$ &  & supersonic  &  and Ar) \\
\hline
$\beta$  & $\gg 1$ near source &  & $\sim 0.35$ (for Ne \\
($P_{\rm pl}/P_{\rm magn}$) & $\ll 1$ at 10 AU  &  &  and Ar) \\
\hline
density contrast   & $>1$  &  & $1-10$ \\ 
($n_{\rm jet}/n_{\rm amb}$) &  &  & \\ 
\hline
\end{tabular} 
\label{ta1}
\end{table}

First of all, the influence of ambient gas was investigated. For this purpose 
the experiments with different variants of gas-puff injection  were performed on 
facilities PF-1000 (Institute of Plasma Physics and Laser Microfusion, Warsaw, 
Poland) and KPF-4 "Phoenix" (Sukhum Physical Technical Institute, Sukhum, Abkhazia). 
On the PF-1000, after the initial filling the chamber  with deuterium (0.9 Torr), 
there was applied an additional injection of a chosen gas (deuterium, helium, neon 
or their mixtures) along the electrode-axis, which influenced on the pinching, 
generation of plasma streams and their propagation through the ambient gas\cite{15, 21}. 
The plasma density was estimated from the Stark broadening at a distance of (27-57) cm 
from the end of the anode, which amounted to $(0.4-4.0) \times 10^{17}$ cm$^{-3}$~\cite{21}, 
which is close to the values obtained on the PF-3 facility. The electron temperature was 
estimated as $3-5$ eV. At modes with gas-puffing, compact plasma structures were 
formed with dimensions of several cm (Fig. 5). Signals from magnetic probes showed 
that inside those plasma structures there were flowing some axial currents~\cite{15}. 
A complicated form of the plasma jet front might be caused by return currents 
which could flow at a plasma stream periphery. 

\begin{figure}[pb]
\centerline{\epsfig{file=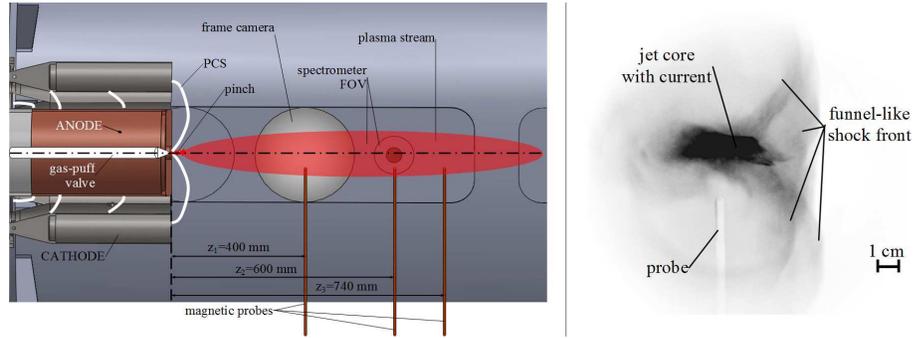,width=12cm}}
\vspace*{8pt}
\caption{a) Scheme of the PF-1000U experiment with the use of the additional gas 
injection.  b) Frame picture of a plasma flow at a distance of 40 cm from the 
anode outlet at the initial deuterium pressure of 1.2 hPa and  the additional 
injection of a mixture of deuterium (75\%) and neon (25\%)\cite{15}.}
\label{fig5}
\end{figure}

In the KPF-4 device we realize the operational mode with the pulsed injection of the 
working gas with increased gas density near the insulator surface and low gas pressure 
in the region of the plasma jet propagation\cite{15}. Thus, the necessary conditions are 
provided for the qualitative development of the discharge, on the one hand, and the 
possibility to change the "contrast" in the flow propagation area by changing the delay 
between the opening of the valve and the initiation of the discharge, on the other 
hand\cite{22}. Note that in this case it was possible to simulate decreasing the density 
of the ambient plasma with a distance from the "central engine", which takes place in 
astrophysical sources. The experiments showed that in such a case, in contrary to 
that when the chamber is filled up to the stationary pressure, the plasma stream 
can propagate without considerable slowing down. In addition, measurements with 
magnetic probes indicate that the magnetic field is concentrated mainly in the 
region with weak luminosity in the optical range, in the so-called "magnetic 
bubbles"\cite{23}. 

An important direction that can be developed in the scheme of the plasma-focus 
experiment is the study of mechanisms for stabilizing the plasma flow, allowing 
it to spread over long distances while maintaining compact transverse dimensions. 
First of all, the analysis of the results obtained in experiments with various 
gases shows the significant effect of the radiation cooling. Compact plasma 
formations at a large distance from the flow generation area were obtained 
in experiments with argon or neon at PF-3 facility (Fig. 4), either with 
additional neon injection into the pinching area on the PF-1000 facility (Fig. 5).  

\begin{figure}[pb]
\centerline{\epsfig{file=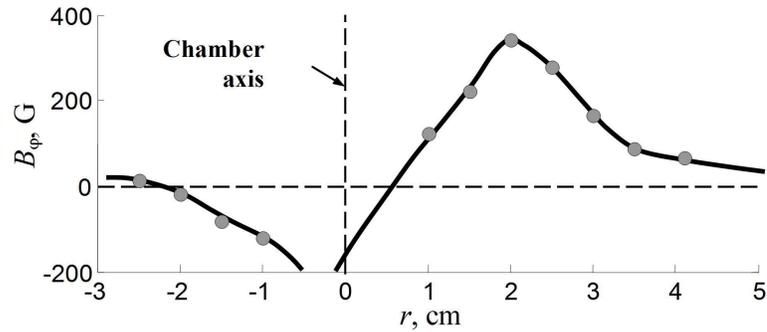,width=10cm}}
\vspace*{8pt}
\caption{Radial distribution of the azimuthal magnetic field in the plasma jet on 
PF-3 facility at a height of $z=35$ cm from the anode plane.}
\label{fig6}
\end{figure}

Finally, another significant result is the fact that the observed radial distribution 
of the azimuthal magnetic field corresponds well to the longitudinal current of $\sim 10$ kA 
flowing in the axial zone, in the core of the flow\cite{13, 20, 24}. It is shown that 
the field distribution has the form $B \propto r$ in the region $r < R_{\rm core}$ and $B \sim 1/r$ 
in the region $r > R_{\rm core}$ (Fig.\ref{fig6}), which is well consistent with the 
known theoretical models. Two important conclusions can be drawn from this. First, the 
estimates show that the magnetic field created by this current can be sufficient to ensure 
the Bennett equilibrium of the plasma. In this case, the stable-state duration of the 
jet should be determined by the time of decay of the currents circulating in the 
plasma. Second, if the axial current is present, it should be closed on the periphery. 

In particular, from the results of experiments at the PF-3 facility it follows that 
the radius of the reverse current  at a distance of 35 cm from the anode is $< 10$ cm [24]. 
The study of the distribution of the reverse current depending on the parameters of 
the experiment is one of the main immediate tasks. 

\section{Conclusion}

Thus, it was shown that the plasma focus in the stage of the pinch formation creates 
a plasma flow propagating along the facility axis with the velocities large than 100 km/s, 
i.e., with velocities close to the one observed in the real jets from young stars. 
It is important that the flow is supersonic as in real jets. The values of the main 
dimensionless parameters achieved in the experiments allow us to consider the PF-type 
facilities as an effective tool for simulation physical processes in the jets of young 
stars. We obtain the conditions when the plasma flow remains compact propagating over 
distances far exceeding its transverse size. This fact indicates the excess of the 
longitudinal jet velocity over the transverse velocity of its expansion. This opens 
the possibility of modeling and studying the mechanisms of stabilization. 

It also was found that many other properties of  the experimental flows  corresponds 
to the main characteristics of young stars jets, for example, the presence of a 
longitudinal current determining the predominance of the toroidal magnetic field. 
However, the nature of the magnetic field distribution, in particular, the detected 
rotation of the induction vector\cite{19, 20}, in the conditions of  the magnetic field 
frozen in the plasma may indicate both the rotation of the flow as a whole, and the 
presence of spiral current structures. Therefore, the question of rotation of the 
plasma flows requires additional studies.  

\section*{Acknowledgments} 

This work is supported by the Russian Science Foundation (project No. 16-12-10051). 
The authors express their appreciation to the PF-3, KPF-4 and PF-1000 teams, 
Yu. G. Kalinin, S. A. Dan'ko and S. S. Anan'ev (NRC Kurchatov Institute, Moscow),  
K. N. Mitrofanov (TRINITI, Troitsk, Moscow) due to efforts of which all experimental 
results were obtained, Ya. N. Istomin (Lebedev Institute, Moscow, Russia) and 
D. L. Grekov (NSC Kharkov Institute of Physics and Technology, Kharkov, Ukraine) 
for fruitful discussions. 

{}

\end{document}